\documentclass[final]{cvpr}

\usepackage{times}
\usepackage{epsfig}
\usepackage{graphicx}
\usepackage{amsmath}
\usepackage{amssymb}
\usepackage{subfigure}
\usepackage{overpic}

\usepackage{bm}
\usepackage{algorithm}

\usepackage{algorithmicx}
\usepackage{algpseudocode}

\usepackage{enumitem}
\setenumerate[1]{itemsep=0pt,partopsep=0pt,parsep=\parskip,topsep=5pt}
\setitemize[1]{itemsep=0pt,partopsep=0pt,parsep=\parskip,topsep=5pt}
\setdescription{itemsep=0pt,partopsep=0pt,parsep=\parskip,topsep=5pt}

\usepackage[pagebackref=true,breaklinks=true,colorlinks,bookmarks=false]{hyperref}

\def\cvprPaperID{159} 
\def\confYear{CVPR 2020}

\graphicspath{{figures/}}

\newcommand{\Ignore}[1]{}

\newcommand{\NullFigure}[3]{}

\title{A Virtual Point Light Generation Method in Close-Range Area}

\author{papers 0578}
\author{Shihao Jin ~~~~~ Rui Wang ~~~~~ Wenting Zheng ~~~~~ Wei Hua ~~~~~ Yuchi Huo
}

\begin{document}

\maketitle

\begin{abstract}
This paper proposes a new hybrid algorithm for sampling virtual point light (VPL). The indirect lighting calculation of the scene is used to distribute the VPL reasonably. In the process of generating VPL, we divide the scene into two parts according to the camera position and orientation. The close-range part: the part that the camera pays attention to. The distant-range part: the part that the camera does not pay attention to or rarely pays attention to. For the close-range part, we use a patch-based vPL sampling method to distribute the VPL as evenly as possible on the patch in the near-field area; for the distant-range part, we use sparse instant radiosity (IR) for sampling. It turns out that, in contrast to conventional multiple instant radiance Compared with the VPL generation algorithm, the method proposed in this paper can greatly improve the quality of the final result graph when the number of VPLs is the same; Under the same rendering quality, the rendering speed can be greatly improved.
\end{abstract}

\section{Introduction}

With the substantial improvement of hardware performance in recent years, industrial applications such as game development, film production, and architectural visualization have higher and higher requirements for realistic Global Illumination (GI) rendering. However, many current algorithms cannot meet the industry's demand for rendering speed and quality requirements. In global illumination rendering, indirect lighting needs to consider the mutual lighting effects of multiple objects in the scene, which is very challenging. Over the years, various methods for calculating indirect lighting have been proposed. Among them, the instant radiance algorithm \cite{r1} generates particles by quasi-random walk method to simulate the diffuse radiance in the scene, which is a very widely used method. For diffuse scenes, the instant radiance algorithm is a very efficient rendering method for indirect lighting.

However, when the camera only focuses on the close-range part of the scene, most of the VPLs generated by the radiance understanding inherited anomalous radiance are located in regions that contribute little or no to the final result image. Scholars have successively proposed some algorithms to improve the distribution of VPL. Among them, the representative algorithms are Metropolis VI)L sampling algorithm \cite{r2}, rejection sampling 131 and multiple vPL generation methods based on bidirectional path tracking \cite{r8,r9}. Compared with these methods, our method inherits the important assumption from the above methods, that is, the area (close-range) observed by the camera needs to be heavily considered. On this basis, it is further proposed that VPL as a light source for indirect illumination calculation needs to be able to provide indirect illumination uniformly, so as to reduce the error caused by uneven distribution of sparse VPL.

Based on the above assumptions, this paper proposes a hybrid algorithm based on the instant radiosity algorithm. First, we divide the scene into close-range views, which the camera mainly focuses on, and distant views, which focus less on them. For close-range parts, we use a patch-based method of sampling virtual point lights; for distant parts, we use a sparsely sampled real-time radiance algorithm. We observe that when the camera only focuses on a small part of the scene, the VPL located in this part contributes more to the final result map than other areas, so a more uniform VPL distribution on the patch is required; In some areas, the VPL in this part has a low contribution to the final result map, so a sparser VPL distribution can be used. Based on the above observations, this paper proposes a hybrid generation method of virtual point light sources that focuses on the mutual illumination of close-range scenes. This method distributes a relatively uniform VPL on the patch for the area that the camera pays attention to, and uses a more uniform VPL for the area that the camera does not pay attention to. Sparse sampling. From the results, compared with the instant radiance algorithm, the rejection sampling algorithm and the Metropolis sampling algorithm, our method can well improve the indirect illumination calculation error caused by the uneven distribution of VPL. Under the same number of VPL, a better rendering quality and faster speed is obtained.

\section{Related Work}

The computation of global illumination relies on computational rendering equations \cite{r4}. Kajiya \cite{r4} proposed a ray-tracing method to compute the rendering equation: tracing the light path backwards from the viewpoint to sample the light reaching the imaging plane. On this basis, scholars have proposed a variety of methods, among which the more representative Monte Carlo ray tracing algorithm \cite{r5} and bilateral path tracing algorithm \cite{r6}. The Monte Carlo ray tracing algorithm is an algorithm similar to Kajiya \cite{r4}, but uses a Monte Carlo method to randomly generate light paths. Monte Carlo ray tracing only traces rays from the camera. On this basis, bilateral path tracing \cite{r5} proposes to trace two paths at the same time, one is from the camera and the other is from the light source.

On the basis of the above two methods, scholars have proposed a variety of methods using spatial light correlation, the most famous example of which is photon mapping \cite{r7}. Photon mapping algorithms propagate and store photons through random walks, and then use the density of these photons distributed in a region of space to estimate the irradiance at a point on a patch. The real-time radiosity algorithm \cite{r1} also uses a (quasi) random walk to generate particles, but the real-time radiosity algorithm replaces the indirect radiance field in the photon map with a virtual point light source.

In the real-time radiosity algorithm, the VPL generated by tracing rays from the light source using the basic VPL tracing algorithm may end up in regions that do not contribute to the resulting image. This is especially true in a large scene where the camera only focuses on a small part of the area. In addition, the VPL generated by the basic condition VI tracking algorithm is not uniformly distributed on the patch, and in some cases, it will bring a large error to the indirect lighting calculation. In order to solve these problems and improve the efficiency of real-time radiometric algorithms, a number of techniques have been proposed to try changing the distribution of VPL.

The simplest of these is rejection sampling \cite{r3}, which rejects those VPLs that do not contribute to the results. In addition to this, there are some more advanced sampling algorithms, such as Metropolis sampling \cite{r2}. However, there exist algorithms \cite{r8,r9} that rely on the tracking produced by the camera rather than the light source path to distribute VPLs.

The main idea of rejection sampling \cite{r3} is simple: they use the same VPL path tracing algorithm as in the instant radiosity algorithm, but reject those VPLs that contribute less than the average value to the resulting image. vPL rejection sampling is simple, effortless, and useful. But this method does not change the distribution function of the PL of the region that contributes greatly to the image. The Metropolis sampling algorithm distributes VPLs according to Metropolis-Hastings sampling. Distribute VPLs in those important areas as much as possible. However, the Metropolis method to obtain a good VPL distribution requires a complex Monte Carlo Markov random process, which requires a lot of time and a large number of random sampling points, so the cost of generating VPL by the Metropolis method is relatively high. Segovia et al. \cite{r8} propose a bidirectional VPL sampling method that simultaneously samples VPLs from light sources and cameras, and then resamples these VPLs. The method proposed by Davidovic \cite{r9} is very related to Segovia's method \cite{r8}, except that the algorithm divides the light transmission into a global part and a local part, and the algorithm is also suitable for glossy scenes. Compared with these methods, the method in this paper can provide a uniform indirect lighting effect by uniformly distributing VI on the patch in the near field, and reduce the error caused by the uneven distribution of sparse VPL.

\section{Close-Range VPL Sampling}
\begin{figure*}
    \centering
    \includegraphics[width=0.9\textwidth]{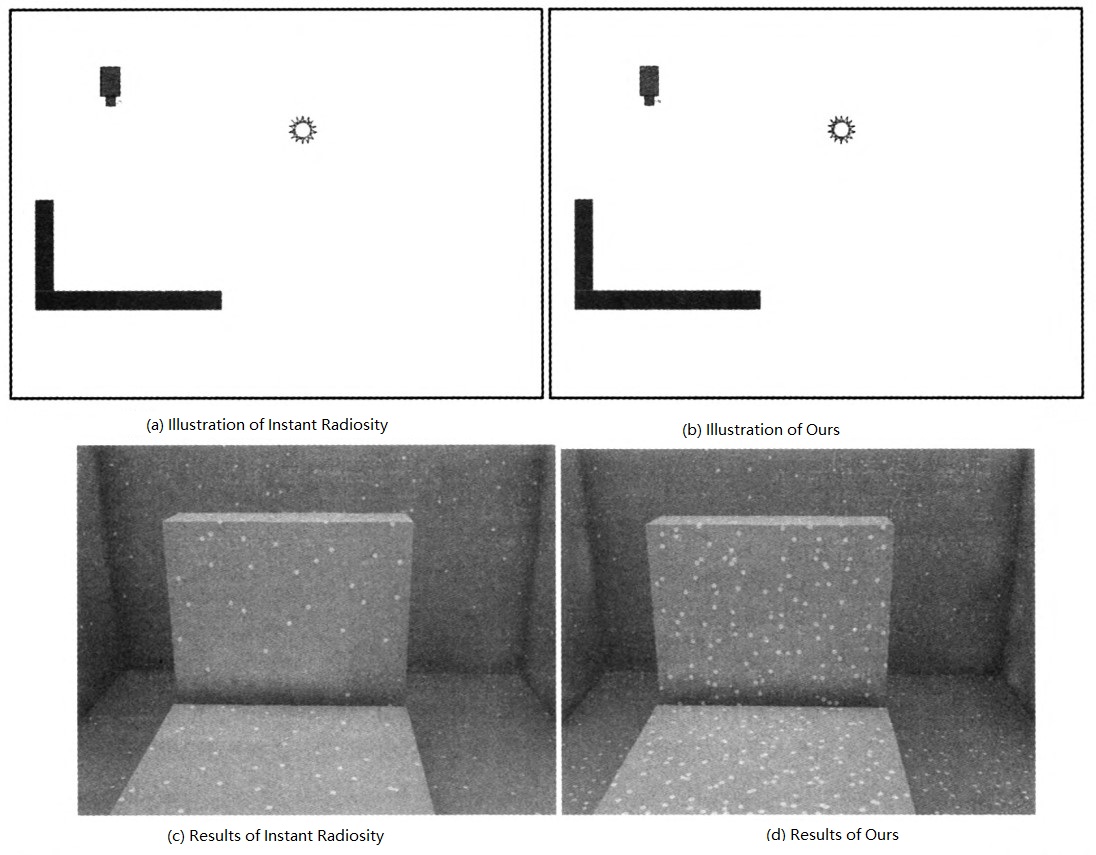}
    \caption{The camera only focuses on a small part of the scene, but the VPLs generated by instant radiance lie mostly outside the region of interest, and these VPLs contribute very little to the final result. However, our method can make most of the VPLs distributed within the region of interest. In the sampling result graph, the dot indicates the position of the VPL. (c) and (d) are the sampling results of the instant radiosity algorithm and our algorithm when the number of VPLs is 1k, respectively.}
    \label{fig1}
\end{figure*}

\textbf{OVerview.} We observe that when the camera only focuses on a part of the scene, most of the VPLs generated by instant radiance are located in regions with little or no contribution to the final result, and are very unevenly distributed in the region of interest, as shown in Figure \ref{fig1}. Show. In order to make VPL more concentrated in the area that contributes to the final result and can provide uniform indirect illumination, we propose a hybrid VPL sampling algorithm for close-range areas. Our algorithm divides the scene into regions that contribute to the resulting map, i.e., the near-field portion, and regions that do not contribute to the resulting image, i.e., the distant-range portion. We use a finer sampling algorithm, patch-based sampling, for close scenes, and a sparser sampling algorithm, immediate radiance with a lower number of particles emitted, for distant scenes. In this way, when the camera only focuses on a small area, our method can make the VPL more concentrated in the contributing area and evenly distributed on the patch.

What we propose is a method to improve the VPL distribution, the basic framework still follows the instant radiosity algorithm. The plotting equation we use is still the plotting equation for instant radiance.

\begin{equation}
    L(y)\approx L_e(y)+\sum_{i=0}^{M-1}G_i(x)V_i(x)L_i,
\end{equation}
where $L_e(y)$ represents direct illumination, $M$ is the number of indirect VPL, $L_i$ is the intensity of VPL $i$, and $G_i$, $V_i$ are its geometry and visibility term, respectively.

In our algorithm, we first distinguish between near and distant views. Here our approach is: for any one-sided face in the scene, if the face is within the camera's field of view, and the distance between the face and the camera is less than a given threshold, we judge that the face is a close-range view; otherwise, it belongs to a long-range view.

The overall flow of the algorithm is as follows: (1) Calculate the close-range part of the patch according to the camera position and orientation; (2) Use the patch-based sampling algorithm to sample the close-range part of the scene; (3) Use the real-time radiometric method to sample the entire scene. Sampling, if the sampling point is in the distant part, the VPL is generated, otherwise the sampling point is discarded. (4) Render the scene with the generated VPLs.

\textbf{VPL Sampling based on patch.} For close-range scenes, we use a patch-based VPL sampling algorithm. The algorithm generates VPLs on each patch visible to the light source.

\textbf{The principle of VPL generation algorithm based on patch.} 
For a patch $i$ in the scene, the estimated value of the vPL generated at the patch center $x$ is:
\begin{equation}
    L_i(x)=G_i(x)V_i(x)L_e,
    \label{eq2}
\end{equation}
where $L_e$ is the light source's intensity, $V_i(x)$ is the visibility term, $G_i(x)=\frac{D\cos{(\omega_i)}}{4\pi r^2}$ is the geometry term, $D$ is the area of patch $i$, $\omega_i$ is the angle between the normal of $i$ and the angle from $x$ toward the camera, $r$ is the distance bwteeen $x$ and the camera.

\textbf{Realization of VPL Generation Algorithm Based on Patch.} 

In the process of implementation, we noticed that the size of the patches in the scene is not very uniform, that is, some patches have a small area, but some patches have a large area. In this case if we still generate a VPL for each patch visible to the light source, then the distribution of VI bars will be very uneven, and if we have a lot of patches, but the number of VPLs we want to generate is much smaller than the number of faces When the number of slices, this generation method is not applicable.

Based on the above observations, we made appropriate adjustments to this algorithm during implementation. First, we do not do VPL generation calculations for all patches. We first use the camera position and perspective to find all the patches in the close-range section that are visible to the light source. Then, for a patch with a larger area, we will generate multiple VPLs on this patch: for a patch with a smaller area, we can use multiple patches with a small area to generate a VPL. Determining which triangles to generate VPL is based on the following method: first sort all triangles (in order to make the variance of the 
sampling result smaller, we sort the triangles in spatial order, so as to ensure that the triangles in the triangle index range are together. We use The spatial sorting method is Morton code \cite{r11}), and the area of each triangle i is denoted as S, then the function of the sum of triangle areas can be expressed as $f(i)=\sum_{j=1}^{i}s_i$, representing the area sum of the first $i$ triangles. Similarly, $f(N)$ represents the  area sum of $N$ triangles in total. We use $g(x)$ to represent the inverse function of $f^{-1}(x)$, which means that if we know an area and $x$, then we can use $g(x)$ to find the subscript $m$ of the triangle index such that $f(m)\leq s<f(m+1)$. 

 This method is to uniformly sample the triangular grid of the close-range area, that is, uniform sampling in the interval $[0,f(N)]$, and the area represented by each VPL is uniform. If $K$ points are sampled, then $[0,f(N)]$ is divided into $K$ parts, $[0, f(N)/K],...,[(K-1)f(N)/K,f(N)]$, then you can find the subscript interval $[1, i_i),...,[i_{k-1},N]$ of the triangle index for each copy, and these triangles are the set of triangles corresponding to a sampling point. Using the rendering equation \ref{eq2}, we use this set of triangles to calculate the brightness of this VPL. When a facet corresponds to multiple VPLs, we sample multiple points on a facet. We randomly generate multiple points \cite{r10} inside the triangle, each point corresponds to a sampling point, and the brightness of each VPL is obtained by bringing the triangle area represented by each sampling point into formula \ref{eq2}.

\subsection{Results}
We analyze the performance of our method by comparing the results with the instant radiosity algorithm, the rejection sampling algorithm, and the Metropolis sampling algorithm. We compared the difference between the results of the two algorithms in the case of equal samples, and the time-consuming of the four algorithms in the case of equal errors and the number of VPLs. Among them, we use the root mean square error (root mean square error, MSE) to measure the error. We use a densely sampled on-the-fly radiosity method (number of VPLs above 200k) to generate reference maps. All statistical data are based on indirect lighting result graphs.
 
 We used a total of three scenarios for testing. Stone carvings and tea sets are a scene with a relatively high number of facets, a desk is an indoor scene with a high facet count and relatively more mutual reflections, and a display cabinet is a scene with richer foreground content. In these scenes, the light sources are all point lights source.
 
\begin{figure*}
    \centering
    \includegraphics[width=0.9\textwidth]{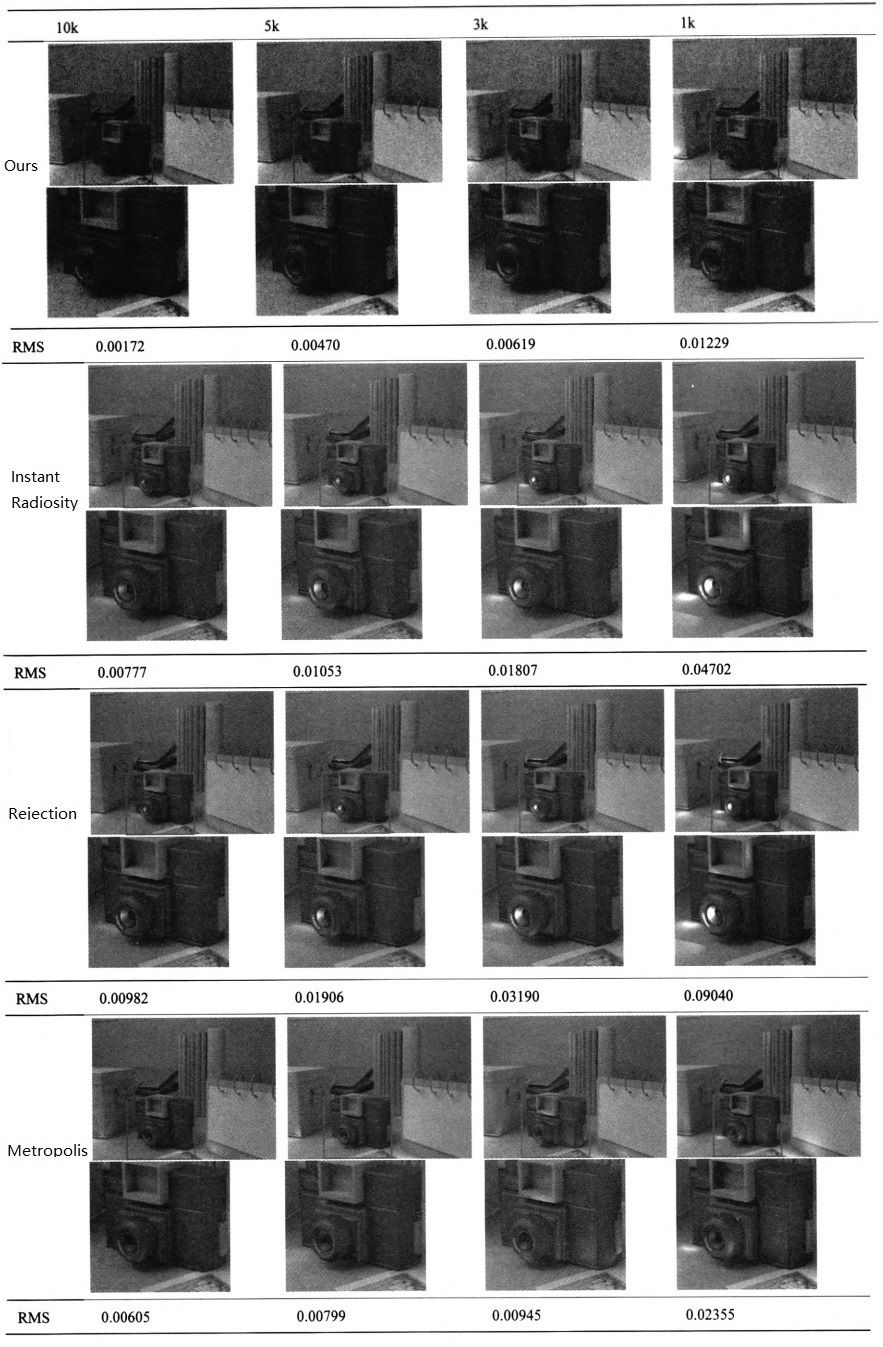}
    \caption{Results comparison. This set of test scenes is a desk scene. We selected the part of the camera with relatively many reflections and complexities for detailed comparison.}
    \label{fig2}
\end{figure*}

 Figure \ref{fig2} shows the test results of the desk scene under the condition of equal samples. Comparing the RMS value and the result graph, we can see that our results are more accurate than the results of the other three algorithms in the case of equal samples, and there are not many false highlights caused by uneven VPL distribution. For ours, the lighting effect in the fruit map is much smoother than that of the instant radiance algorithm and the rejection sampling algorithm. When the sample size is 5k, these two methods have obvious error highlighting, but our method has reduced the sample size to 5k. lk, can still maintain a good effect. Although the results of the Metropolis sampling algorithm only show clearly false highlights when the samples are down to 1k, the RMS of our method is about half that of the method in the equal sample case.
 
\begin{figure*}
    \centering
    \includegraphics[width=0.9\textwidth]{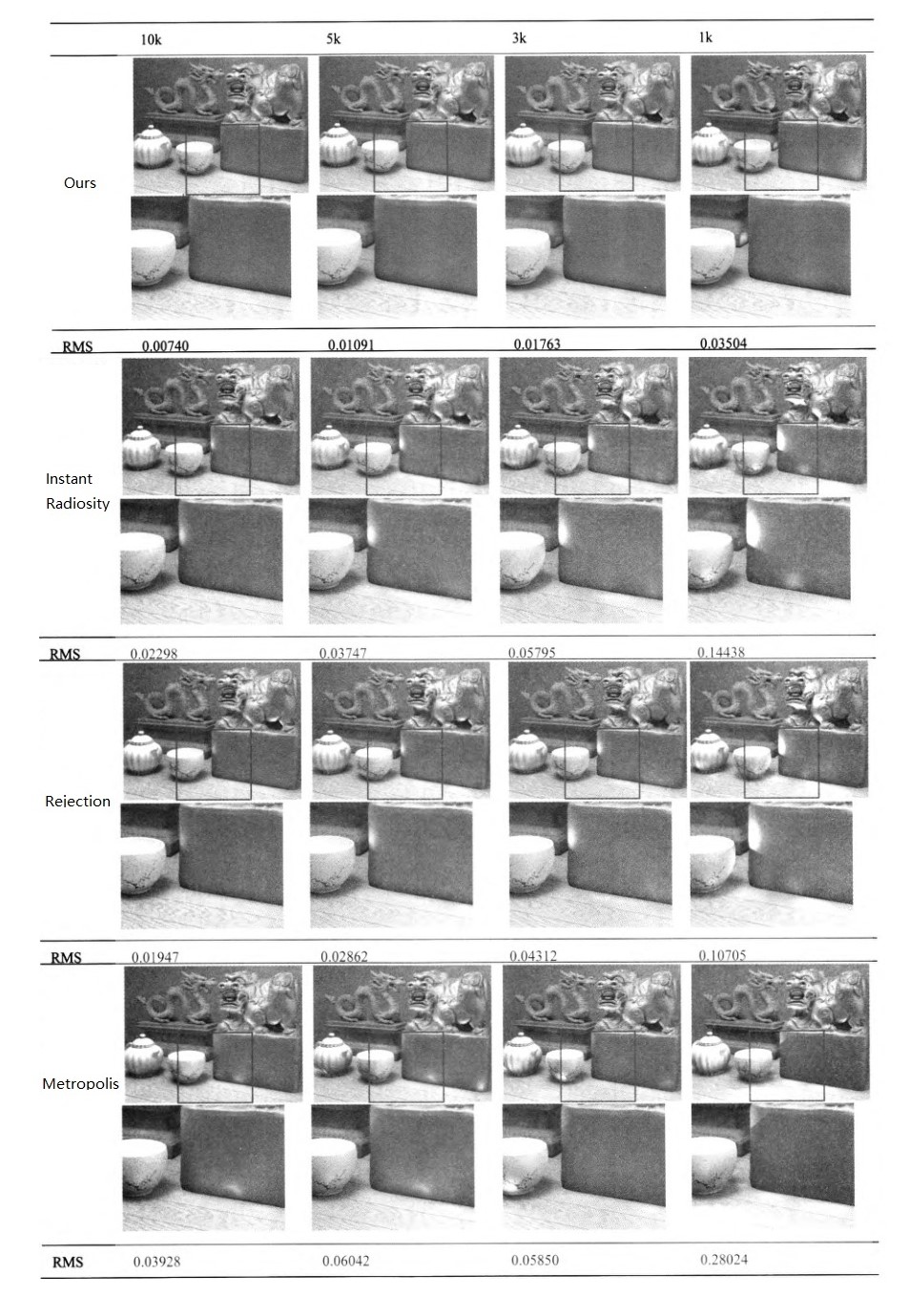}
    \caption{Results comparison. This set of test scenes are stone carvings and tea sets. We select the lower part of the stone lion for detailed comparison.}
    \label{fig3}
\end{figure*}

 Figure \ref{fig3} shows the test results of the stone carving and the tea set scene under the same sample condition. By comparing the results, it can be seen that our method significantly outperforms the other three methods. In the case of equal samples, the RMS of the other three methods is about more than 3 times that of our method. When the number of samples is approaching 1k, our method still works well, while the other three methods have obvious highlight errors at 5k.

\begin{figure*}
    \centering
    \includegraphics[width=0.9\textwidth]{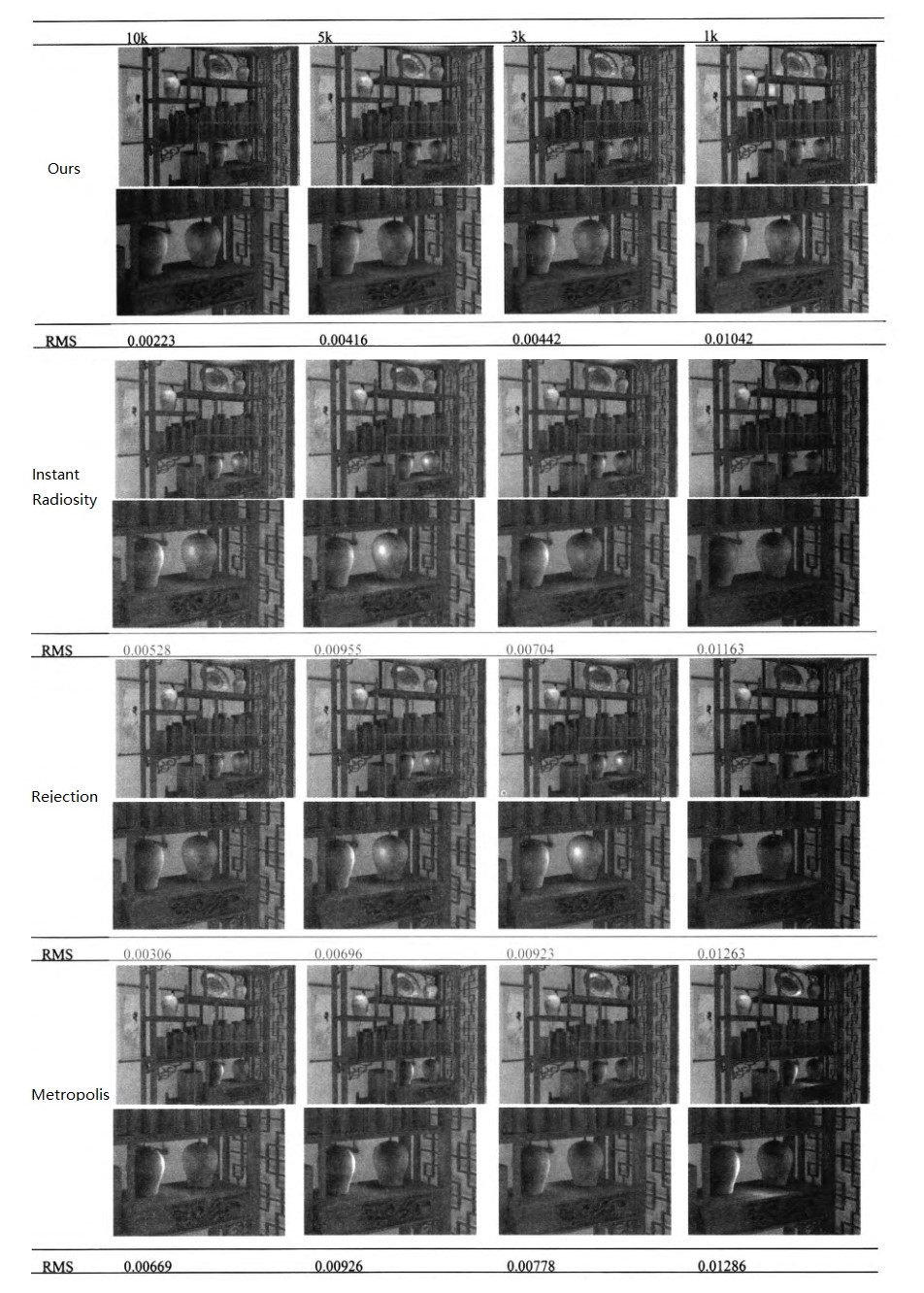}
    \caption{Results comparison. This set of test scenarios is the showcase scenario. We select the part of the crock below the display cabinet for detailed comparison.}
    \label{fig4}
\end{figure*}

 Figure \ref{fig4} shows the test results of the showcase scenario under the condition of equal samples. For the showcase scene, when the number of samples is larger than lk, our method significantly outperforms the instant radiosity algorithm and the Metropolis sampling algorithm, and the RMS of our method is only about half of the latter two. Although the results of the rejection sampling algorithm are also good, there are obvious errors highlighted in the results of this method between the crocks under the display case.

\begin{figure*}
    \centering
    \includegraphics[width=1.0\textwidth]{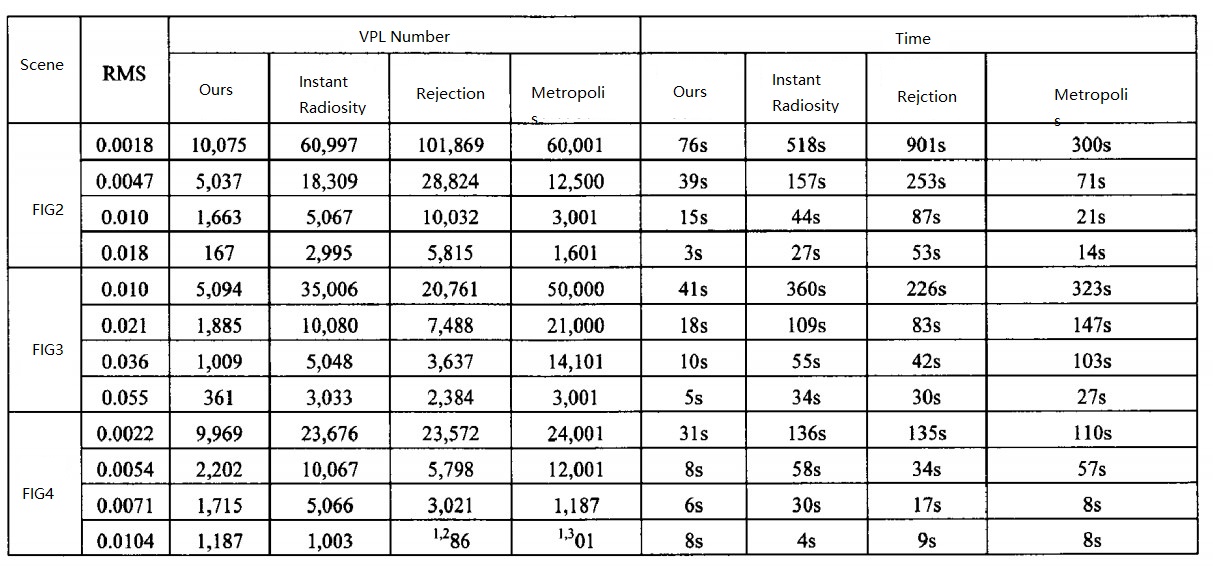}
    \caption{Equal RMSE test. }
    \label{tab1}
\end{figure*}

Figure \ref{tab1} shows the test results of the three scenarios under the condition of equal RMSE. It can be seen from the table] that for the two scenes of desk and stone carving and tea set, in the case of waiting for RMSE, the number of vPLs of the other three methods is 2-5 times of the number of VPLs of our algorithm, and their running time is also our algorithm. 2-5 times the running time of the method. For the showcase scenario, when P, MS is relatively low, eg RMS is 0.0022 and O. 0054, our method has obvious advantages, and the number of VPLs and running time of the other three methods are more than 2 times that of our method. When the RMS is relatively low, for example, the RMS is O. 0104, the effect of these four methods is not very different. Overall, our algorithm performs significantly better than the other three methods.

\section{Conclusion}
This paper proposes a new VPL sampling method. For the case where the camera only focuses on a part of the scene, we design a hybrid VPL sampling method, which distributes the sampling points evenly on the patch in the near-field area, and uses the sparse instant radiometric method for sampling in the distant-range area. The results show that, compared with the instant radiosity algorithm, the rejection sampling algorithm and the Metropolis sampling algorithm, the method proposed in this paper can obtain better rendering effect under the same number of VPLs, and obtain faster rendering speed under the same rendering quality.

There are some extended articles about applying physical lighting computation in various applications:

\begin{enumerate}
    \item Deep Learning-Based Monte Carlo Noise Reduction By training a neural network denoiser through offline learning, it can filter noisy Monte Carlo rendering results into high-quality smooth output, greatly improving physics-based Availability of rendering techniques \cite{huo2021survey}, common research includes predicting a filtering kernel based on g-buffer \cite{bako2017kernel}, using GAN to generate more realistic filtering results \cite{xu2019adversarial}, and analyzing path space features Perform manifold contrastive learning to enhance the rendering effect of reflections \cite{cho2021weakly}, use weight sharing to quickly predict the rendering kernel to speed up reconstruction \cite{fan2021real}, filter and reconstruct high-dimensional incident radiation fields for unbiased reconstruction rendering guide \cite{huo2020adaptive}, etc.
    \item The multi-light rendering framework is an important rendering framework outside the path tracing algorithm. Its basic idea is to simplify the simulation of the complete light path illumination transmission after multiple refraction and reflection to calculate the direct illumination from many virtual light sources, and provide a unified Mathematical framework to speed up this operation \cite{dachsbacher2014scalable}, including how to efficiently process virtual point lights and geometric data in external memory \cite{wang2013gpu}, how to efficiently integrate virtual point lights using sparse matrices and compressed sensing \cite{huo2015matrix}, and how to handle virtual line light data in translucent media \cite{huo2016adaptive}, use spherical Gaussian virtual point lights to approximate indirect reflections on glossy surfaces \cite{huo2020spherical}, and more.
    \item Automatic optimization of rendering pipelines Apply high-quality rendering technology to real-time rendering applications by optimizing rendering pipelines. The research contents include automatic optimization based on quality and speed \cite{wang2014automatic}, automatic optimization for energy saving \cite{ wang2016real,zhang2021powernet}, LOD optimization for terrain data \cite{li2021multi}, automatic optimization and fitting of pipeline rendering signals \cite{li2020automatic}, anti-aliasing \cite{zhong2022morphological}, etc.
    \item Using physically-based process to guide the generation of data for single image reflection removal \cite{kim2020single}; propagating local image features in a hypergraph for image retreival \cite{an2021hypergraph}; managing 3D assets in a block chain-based distributed system \cite{park2021meshchain}.
\end{enumerate}

\bibliographystyle{ieee}
\bibliography{srbib}

\end{document}